\documentclass[fp,twocolumn,seceq,dvipdfmx]{jpsj3}
\usepackage{graphicx}
\def\v#1{\mib #1}

\newcommand{\bra}[1]{\left\langle {#1} \right\vert}
\newcommand{\ket}[1]{\left\vert {#1} \right\rangle}

\def\Jad{J_{\rm ad}}

\def\Jeffa{J_{\rm eff}^{\rm A}}
\def\Jeffb{J_{\rm eff}^{\rm B}}
\def\Ns{N_{\rm s}}
\def\H{{\cal H}}
\def\Heffa{{\cal H}_{\rm eff}^{\rm A}}
\def\Heffb{{\cal H}_{\rm eff}^{\rm B}}
\def\deltaa{{\delta_{\rm A}}}

\def\deltab{{\delta_{\rm B}}}

\def\theenumi{{\roman{enumi}}} 

\title
{
Ground State Phases of Distorted $S=1$ Diamond Chains}
\author
{
Kazuo  Hida\thanks{E-mail address: hida@mail.saitama-u.ac.jp}
}

\inst
{Professor Emeritus, Division of Material Science, Graduate School of Science and Engineering, \\ 
Saitama University, Saitama, Saitama, 338-8570\\ 
}

\recdate
{
\today
}

\abst
{
The ground states of  distorted $S=1$ diamond chains are investigated for two types of distortion called type A and B [J. Phys. Soc. Jpn. 79 (2010) 114703]. 
For the type A distortion, Haldane phases with and without spontaneous translational symmetry  breakdown are present for large values of parameter $\lambda$ that parametrize the strength of frustration. For small $\lambda$, the Haldane phase and two quantized ferrimagnetic phases in the undistorted chain remain stable even for strong distortion. 
In contrast, for the type B distortion,  the quantized ferrimagnetic phases  with and without spontaneous translational symmetry  breakdown are present for large $\lambda$. The partial ferrimagnetic phases emerge between them. For small $\lambda$, two quantized ferrimagnetic phases remain and the partial ferrimagnetic phases also emerge between them. The Haldane phase between the two kinds of ferrimagnetic phases turns into a topologically trivial double Haldane phase for strong distortion.
}

\begin{document}
\sloppy
\maketitle
\section{Introduction}

Exotic ground state phases of low-dimensional quantum frustrated spin systems have been attracting broad interest in recent condensed matter physics from experimental and theoretical viewpoints.\cite{intfrust,diep} To understand the nature of these phases theoretically, the frustrated spin models with exact ground states are helpful starting points. The Majumdar-Ghosh model\cite{mg} and  the Shastry-Sutherland model\cite{shs} are examples of such models that have ground states consisting of singlet dimers.

\begin{figure} 
%\centerline{\includegraphics[width=6cm]{lattice_s1dia_s.eps}}
\centerline{\includegraphics[width=6cm]{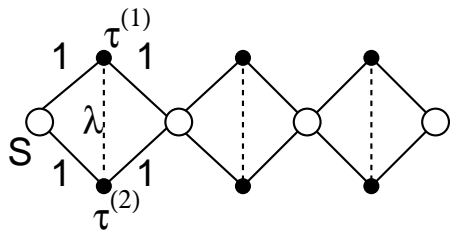}}
\caption{
Structure of the diamond chain. 
Spin magnitudes in a unit cell are indicated by $S$. $\tau^{(1)}$ and $\tau^{(2)}$; 
we denote the set of magnitudes by ($S$, $\tau$) where  $\tau^{(1)}=\tau^{(2)}=\tau$. 
We consider the case $S = \tau=1$ in the present paper.}
\label{lattice_structure}
\end{figure}

The diamond chain discussed in the present work is another frustrated spin chain with exact ground states. 
The lattice structure is shown in Fig.~\ref{lattice_structure}. 
In a unit cell, there are two kinds of nonequivalent lattice sites occupied by spins with magnitudes $S$ and $\tau$; we denote the set of magnitudes by ($S$, $\tau$) where spins with magnitude $S$ are on the vertex sites and those with magnitude $\tau$ are on the apical sites. Two $\tau$-spins are connected by the vertical bond with strength $\lambda$.  
 The features common to all types of diamond chains are their infinite number of local conservation laws and more than two different types of exact ground states that are realized depending on the strength of frustration. 

Takano and coworkers\cite{takano,Takano-K-S} introduced this lattice structure and generally investigated the case of ($S$, $S$).  
Particularly, in the case of (1/2, 1/2), they determined the full phase diagram of the ground state by combining rigorous arguments with numerical calculations. 

%====================================
%====================================

The ground states of spin-1 diamond chains (S1DC) with $(S,\tau)=(1,1)$ are further studied by Hida and Takano\cite{ht_s1}. In the strongly frustrated regime, the ground state of the S1DC is same as that of the mixed diamond chain with  $(S,\tau)=(1,1/2)$.\cite{tsh} Three different paramagnetic phases accompanied by spontaneous translational symmetry breakdown (STSB) and one paramagnetic phase without STSB are found in this regime.  This model also has a nonmagnetic Haldane phase and ferrimagnetic phases  with and without STSB in a less frustrated regime.\cite{ht_s1}  

\begin{figure}[t] 
%\centerline{\includegraphics[width=6cm]{lattice_s1dia_s_stag.eps}}
%\centerline{\includegraphics[width=6cm]{lattice_s1dia_s_uni.eps}}
\centerline{\includegraphics[width=6cm]{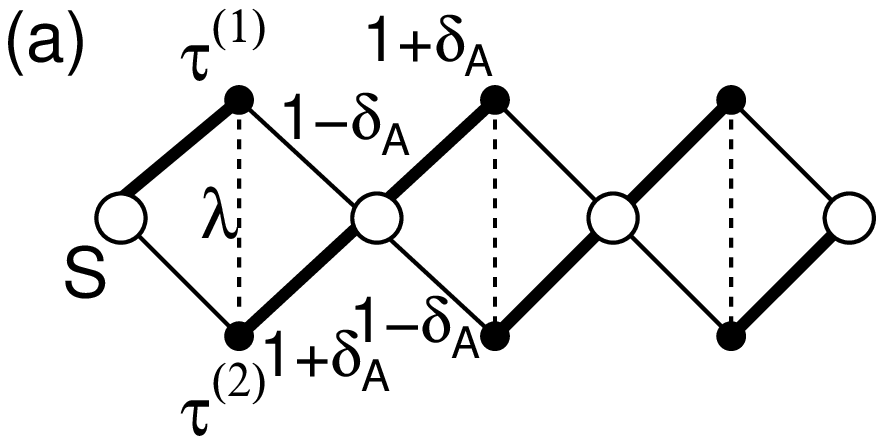}}
\centerline{\includegraphics[width=6cm]{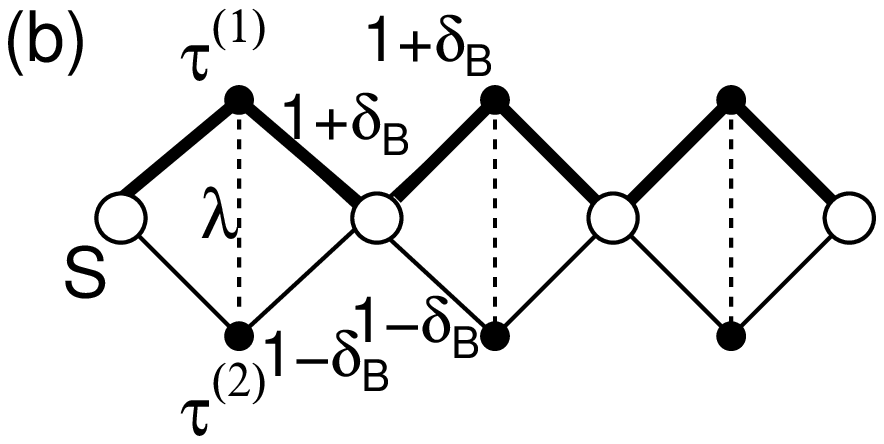}}
\caption{Structures of S1DC  with (a) type A and (b) type B distortions.}
\label{lattice}
\end{figure}

In the present paper, we study the effect of distortion on the ground states of S1DC. We investigate the distortion patterns  depicted in Figs. \ref{lattice}(a) and \ref{lattice}(b).  Following Ref. \citen{hts_distort}, we call the distortion patterns in Fig. \ref{lattice}(a) and Fig. \ref{lattice}(b) as type A and type B, respectively. As discussed in Ref. \citen{hts_distort}, these types of  distortion break the local conservation laws that hold in the undistorted S1DC. As a result, they induce effective interactions between the cluster spins, and form novel exotic phases such as  Haldane phases with STSB, ferrimagnetic phases with STSB, and partial ferrimagnetic (PF) phase that can be regarded as spontaneously magnetized Luttinger liquid\cite{furuya,sekiguchi}. In addition, a double Haldane phase\cite{hiki1,hiki2,kol1,kol2,kol3} is found for strong type B distortion. It should be remarked that the S1DC with type A distortion is realized in real material,\cite{kuni,kiku1} although the exchange parameters of the material cannot be controlled freely to realize the most exotic Haldane phases with STSB. 
%=====================================
%=====================================

This paper is organized as follows. 
In \S 2, the Hamiltonians for the S1DCs with type A and type B distortions are presented, and the structure of the ground states of the S1DC without distortion is summarized. 
The ground-state phases for the S1DC with type A distortion are discussed in \S 3, and those for the S1DC with type B distortion are discussed in \S 4. 
The last  section is devoted to summary and discussion.

\section{Hamiltonian}
\label{section:ham}

The S1DCs with type A and type B distortions are described, respectively, by the following Hamiltonians: 
%------------------------------------------------------------
\begin{align}
{\cal H}_{\rm A}& = \sum_{l=1}^{N} 
\Bigl[ (1+\deltaa)\v{S}_{l}\v{\tau}^{(1)}_{l}
+ (1-\deltaa)\v{\tau}^{(1)}_{l}\v{S}_{l+1}
\nonumber\\
&+(1-\deltaa)\v{S}_{l}\v{\tau}^{(2)}_{l} 
+(1+\deltaa)\v{\tau}^{(2)}_{l}\v{S}_{l+1} 
+ \lambda\v{\tau}^{(1)}_{l}\v{\tau}^{(2)}_{l} \Bigr] , 
\label{hama}\\
{\cal H}_{\rm B}& =\sum_{l=1}^{N} 
\Bigl[ (1+\deltab)\v{S}_{l}\v{\tau}^{(1)}_{l}
+(1+\deltab)\v{\tau}^{(1)}_{l}\v{S}_{l+1}
\nonumber\\
&+(1-\deltab)\v{S}_{l}\v{\tau}^{(2)}_{l}
+(1-\deltab)\v{\tau}^{(2)}_{l}\v{S}_{l+1} 
+ \lambda\v{\tau}^{(1)}_{l}\v{\tau}^{(2)}_{l} \Bigr] , 
\label{hamb}
\end{align}
%------------------------------------------------------------
where $\v{S}_{l}$, $\v{\tau}^{(1)}_{l}$ and $\v{\tau}^{(2)}_{l}$ 
are the spin-1 operators in the $l$th unit cell. 
The parameter $\deltaa$ ($\deltab$) represents the strength of type A (type B) distortion, and is taken to be nonnegative without spoiling generality. 
The number of unit cells is denoted by $N$, 
and then the total number of sites $\Ns$ is $3N$.

For $\deltaa$ = 0 and $\deltab$ = 0, both eqs.~(\ref{hama}) and (\ref{hamb}) reduce to the Hamiltonian of the undistorted S1DC as, 
%------------------------------------------------------------
\begin{align}
{\cal H}_0 &=\sum_{l=1}^{N} \left[\v{S}_{l}\v{T}_{l}+\v{T}_{l}\v{S}_{l+1}+ \frac{\lambda}{2}\left(\v{T}^2_{l}-\frac{3}{2}\right)\right], 
\label{ham2}
\end{align}
%------------------------------------------------------------
where the composite spin operator $\v{T}_l$ is defined by 
$\v{T}_l\equiv\v{\tau}^{(1)}_{l}+\v{\tau}^{(2)}_{l}$. 
Before going into the analysis of the distorted S1DC, we briefly summarize the ground-state properties of the Hamiltonian (\ref{ham2}) reported in Refs. \citen{Takano-K-S} and \citen{ht_s1} for convenience.
\begin{enumerate}
\item 
$\v{T}_l^2$ commutes with the Hamiltonian ${\cal H}_0$ for any $l$. 
Therefore, the composite spin magnitude $T_l$ 
 defined by $\v{T}_l^2 = T_l(T_l+1)$ is a good quantum number that takes the values 0, 1, and 2. 
Hence, each energy eigenstate has a definite set of $\{T_l\}$, i.e. a sequence of 0's, 1's and 2's with length $N$. 
A pair of $\v{\tau}^{(1)}_{l}$ and $\v{\tau}^{(2)}_{l}$ with $T_l=0$ is called a dimer. 
 A cluster including $n$ successive pairs with $T_l \neq 0$ bounded by a pair of dimers is called a cluster-$n$. 

\item For $\lambda >2$, $T_l=2$ is not allowed in the ground state.
\item There are 4 distinct paramagnetic ground-state phases called dimer-cluster-$n$ (DC$n$) phases with $n=0,1,2,3$. The DC$n$ state is an alternating array of dimers and cluster-$n$'s. 
There are also 3 phases with a single infinite cluster that correspond to a Haldane phase (HDC$\infty$ phase: $\forall l\ T_l=1$), a ferrimagnetic phase with $m=1/6$ (F$_{1/6}$ phase; $\forall l\ (T_{2l},T_{2l+1})=(1,2)$ or $(2,1)$, and $m=1/3$ (F$_{1/3}$ phase; $\forall l\ T_l=2$), where $m=M/\Ns$ is the spontaneous magnetization $M$ per site. 
The phase boundary $\lambda_{\rm c}(n,n')$ between DC$n$ and DC$n'$ phases are given by
%------------------------------------------------------------
\begin{align}
&\lambda_{\rm c}(0,1) = 3, \\
&\lambda_{\rm c}(1,2) \simeq 2.6604, \\
&\lambda_{\rm c}(2,3) \simeq 2.5827.
\label{lambdac1}
\end{align}
The DC3-Haldane,  Haldane-F$_{1/6}$, and F$_{1/6}$-F$_{1/3}$ phase boundaries are denoted by $\lambda_{\rm c}(3,{\rm H})$, $\lambda_{\rm c}({\rm H}, {\rm F}_{1/6})$ and $\lambda_{\rm c}({\rm F}_{1/6}, {\rm F}_{1/3})$, respectively. They are given by
\begin{align}
&\lambda_{\rm c}(3,{\rm H}) \simeq 2.5773,\\
&\lambda_{\rm c}({\rm H}, {\rm F}_{1/6}) \simeq 1.0727, \\
&\lambda_{\rm c}({\rm F}_{1/6}, {\rm F}_{1/3}) \simeq 1.0182.
\label{lambdac4}
\end{align}
%------------------------------------------------------------
\item The DC$n$ phases with $0 \leq n \leq 3$ are realized for  $\lambda > \lambda_{\rm c}(3,{\rm H})$. Since $\lambda_{\rm c}(3,{\rm H})>2$, we have $T_l=1$ within each cluster-$n$. This implies that a cluster-$n$ is equivalent to the ground state of an antiferromagnetic spin-1 Heisenberg chain of length $2n+1$ with open boundary condition.
\item In the DC$n$ phase with $1\leq n \leq 3$, $(n+1)$-fold STSB takes place. In the F$_{1/6}$ phase, two-fold STSB takes place.  In the DC$0$, HDC$\infty$, and F$_{1/3}$ phases, no translational symmetry is broken.
\end{enumerate}
In what follows, we examine the effects of the type A and type B distortions on the ground state of S1DC analytically and numerically. Because the DC3 phase is only realized  within a very narrow interval of $\lambda$, numerical analyses are difficult in this  phase. Hence, we do not consider  the DC3 phase in the following numerical analyses.

\section{Ground-State Properties of the S1DC with Type A Distortion}

\subsection{Weak distortion regime $(\deltaa \simeq 0)$}

\begin{figure} 
%\centerline{\includegraphics[width=7cm]{hdcss1ns_gray.eps}}
\centerline{\includegraphics[width=7cm]{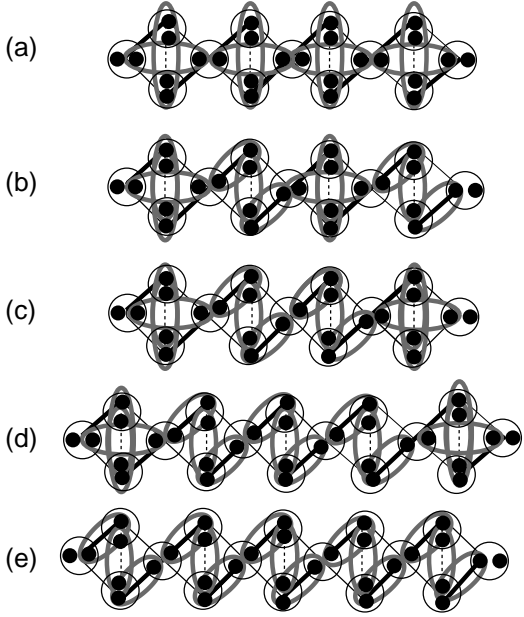}}
\caption{Valence bond structures of the nonmagnetic ground state phases of S1DC with type A distortion for (a) HDC0 (uniform Haldane), (b) HDC1, (c)HDC2, (d) HDC3, and  (e) HDC$\infty$ (uniform Haldane) phases. The big open circles indicate the spin-1 sites. One spin-1 site consists of two spin-1/2's depicted by the filled small circles that are symmetrized within each open circle.  Each gray oval encircles a singlet pair of two spin-1/2's. The thick solid, thin solid, and dotted lines are the bonds with strength $1+\deltaa$, $1-\deltaa$, and $\lambda$, respectively.}
\label{valence}
\end{figure}
For $\lambda > 2$, only the states with $T_l=0$ and 1 are allowed. Hence, the argument proceeds in the same way as the case of $(S,\tau)=(1,1/2)$\cite{hts_distort}. For the type A distortion, the total spins of the cluster-$n$'s on both sides of a 
dimer tend to be antiparallel to each other. Namely, the effective coupling between the spins of neighboring cluster-$n$'s is antiferromagnetic. As in Ref. \citen{hts_distort}, we have estimated the ratio of the effective bilinear and biquadratic interactions between the spins of cluster-$n$ and confirmed that the ground state of the whole chain is a Haldane state. 
We call this state 
the Haldane DC$n$ (HDC$n$) state. The valence bond structures for the HDC$n$ phases with $n=0, 1, 2, 3$ and $\infty$ 
 are shown in Fig. \ref{valence}. 
In the HDC$n$ state with $1 \leq n \leq 3$, the $(n+1)$-fold translational symmetry is spontaneously broken 
 unlike the conventional Haldane state.  On the other hand, the Haldane state in the undistorted S1DC is robust against distortion, since this ground state has an energy gap. 
Both the HDC0 state  for $\lambda > \lambda_{\rm c}(0,1)$ and the Haldane state for $\lambda_{\rm c}({\rm H},{\rm F}_{1/6}) < \lambda < \lambda_{\rm c}(3,{\rm H})$ are the Haldane states without STSB. 
 The ferrimagnetic F$_{1/6}$ and F$_{1/3}$ phases with finite long range orders are also robust against weak distortion. 

\subsection{Strong distortion regime $(\deltaa \simeq 1)$}\label{subsec:stronga}

For $\deltaa = 1$ and  $\lambda=0$, the three spins $\v{\tau}^{(2)}_{l-1}$, $\v{S}_{l}$, and $\v{\tau}^{(1)}_{l}$ form a three spin cluster.  The cluster Hamiltonian is given by
\begin{align}
{\cal H}_3&=2J\left[\v{\tau}^{(2)}_{l-1}\v{S}_{l}+\v{S}_{l}\v{\tau}^{(1)}_{l}\right]\label{eq:ham3}
\end{align}
This can be regarded as a spin-1 antiferromagnetic Heisenberg chain with length 3. According to the Marshal-Lieb-Mattis theorem, the total spin of the ground state of this Hamiltonian is unity. Hence, each cluster carries an effective spin $\tilde{\v{S}}_l(=\v{\tau}^{(2)}_{l-1}+\v{S}_{l}+\v{\tau}^{(1)}_{l}$) with magnitude 1. The three spin ground state $\ket{G;\tilde{S}^z_l}$ with $\tilde{S}^z_l=1$ is expressed using the basis $\ket{{\tau}^{(2)z}_{l-1} \ {S}_{l}^z \  {\tau}^{(1)z}_{l}}$ as
\begin{align}
\ket{G;1}&=\sqrt{\frac{3}{20}}\left[-\frac{1}{3}\left(\ket{11\bar{1}}+\ket{\bar{1}11}\right)\right.\nonumber\\
&\left.-{2}\ket{1\bar{1}1}
+\left(\ket{0 0 1}+\ket{1 0 0 }\right)-\frac{2}{3}\ket{01 0}\right]
\end{align}
where $\bar{1}$ stands for $-1$. The eigenstates $\ket{G;0}$ and $\ket{G;\bar{1}}$ 
 are obtained by applying the descending operator on $\ket{G;1}$. Up to the first order in $\lambda$ and $1-\deltaa$, the effective interaction between $\tilde{\v{S}}_l$s can be described by the Hamiltonian
\begin{align}
\Heffa&=\sum_{l=1}^{N} \Jeffa\tilde{\v{S}}_{l}\tilde{\v{S}}_{l+1},
\end{align}
where
\begin{align}
\Jeffa&=-\frac{3}{4}(1-\deltaa)+\frac{9}{16}\lambda.
\end{align}
Hence, the ground state is the Haldane state 
 consisting of $\tilde{\v{S}}_l$s for $\lambda > \frac{4}{3}(1-\deltaa) $, while it is a ferrimagnetic state with $m=1/3$ (ferromagnetic in terms of $\tilde{\v{S}}_l$) for $\lambda<\frac{4}{3}(1-\deltaa)$. This is consistent with the numerical phase diagram discussed in the next section around $(\lambda,\delta)=(0,1)$ as shown by the dotted line in Fig. \ref{phase_stag}. 

 The Haldane ground state for $\lambda>\frac{4}{3}(1-\deltaa) $ is not accompanied by STSB. This nature is common to the HDC0 and HDC$\infty$ phases in the weak distortion limit. 
Furthermore, the HDC$\infty$ state is transformed into 
 the HDC0 state only by rearranging two valence bonds within each diamond unit, as can be seen from Figs.~\ref{valence}(a) and \ref{valence}(e). 
Also, considering that the Hamiltonian (\ref{eq:ham3}) can be regarded as a spin-1 chain with length 3, the ground state of the 3-spin cluster has the valence bond solid structure.\cite{aklt1,aklt2} Thus the valence bond structure of the Haldane phase consisting of  $\tilde{\v{S}}_l$s is just as depicted in Fig. \ref{valence}(e).
Therefore, the Haldane state consisting of $\tilde{\v{S}}_l$s, HDC0, and HDC$\infty$ should be regarded as different parts of a single phase. 
The continuity of the three regimes will be confirmed by the numerical analysis discussed in \S 3.3. 
In what follows, we call this phase the uniform Haldane (UH) phase as a whole.

The ferrimagnetic ground state for $\lambda<\frac{4}{3}(1-\deltaa)$ has $m=1/3$. This phase is connected to the F$_{1/3}$ phase in the absence of distortion.
\subsection{Numerical phase diagram}

\begin{figure} 
%\centerline{\includegraphics[width=7cm]{critical_stagwhole_0s.eps}}
\centerline{\includegraphics[width=7cm]{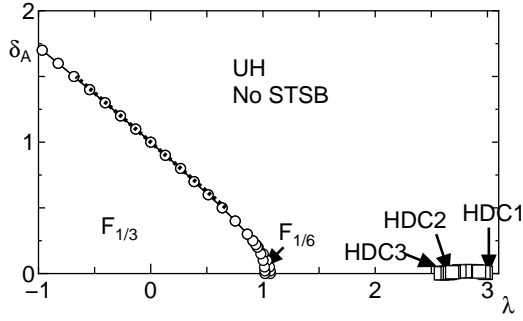}}
\caption{Phase diagram of the S1DC with type A distortion. The dotted line is the phase boundary  $\lambda = \frac{4}{3}(1-\deltaa) $obtained by the strong distortion approximation in sect. \ref{subsec:stronga}. Enlarged figures for HDC phases and F$_{1/6}$ phase are shown in Figs. \ref{phase_stag_hal} and \ref{phase_stag_ferri}, respectively.}
\label{phase_stag}
\end{figure}
\begin{figure} 
%\centerline{\includegraphics[width=7cm]{critical_dm.eps}}
\centerline{\includegraphics[width=7cm]{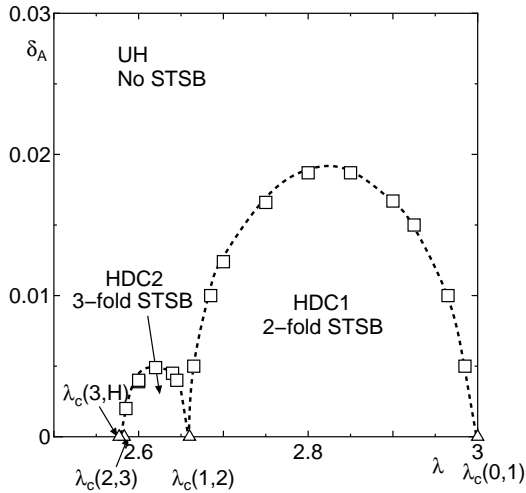}}
\caption{Enlarged phase diagram of the S1DC with type A distortion for strongly frustrated regime (large $\lambda$) determined by the DMRG method. The triangles indicate the position of the phase boundary for $\deltaa=0$.}
\label{phase_stag_hal}
\end{figure}
\begin{figure} 
%\centerline{\includegraphics[width=7cm]{critical_stag_ls.eps}}
\centerline{\includegraphics[width=7cm]{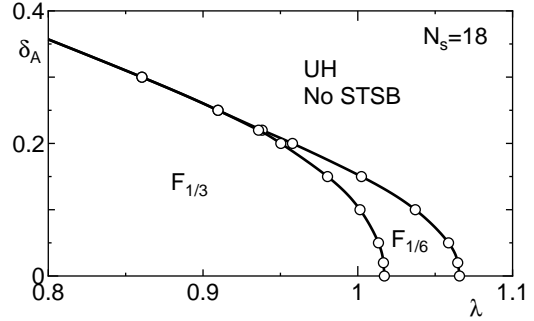}}
\caption{Enlarged phase diagram of the S1DC with type A distortion for weakly frustrated regime (small $\lambda$) determined by numerical diagonalization with $\Ns=18$. 
}
\label{phase_stag_ferri}
\end{figure}
\begin{figure} 
%\centerline{\includegraphics[width=7cm]{mag_st_dlt010s.eps}}
\centerline{\includegraphics[width=7cm]{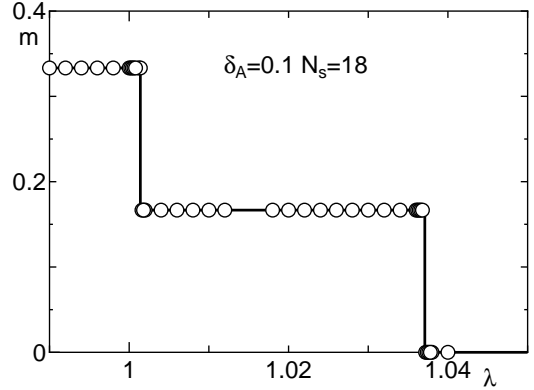}}
\caption{$\lambda$-dependence of the spontaneous magnetization for $\delta_{\rm A}=0.1$ with $\Ns=18$.}
\label{fig:ms_stag}
\end{figure}

Following Ref. \citen{hts_distort}, we employ the DMRG calculation with the open boundary condition to determine the phase diagram for finite $\deltaa$ in the region $\lambda > \lambda_{\rm c}(3,{\rm H})$. In the DMRG calculation, the number of states $\chi$ kept in each subsystem ranged from 200 to 400, and the system size $\Ns$ from 60 to 288.
From the valence bond structures for the HDC$n$ phases in Fig. \ref{valence}, the HDC$n$ ground state has translational invariance of period $n+1$. Hence, the $(n+1)$-fold  STSB  takes place at the HDC$n$-UH phase boundary.  
As in Ref. \citen{hts_distort}, we carried out the size extrapolation of the order parameter assuming  the 2-dimensional $(n+1)$-clock model universality class. 
 The results are shown in the phase diagrams of Figs. \ref{phase_stag} and  \ref{phase_stag_hal}. The error bars are within the size of the symbols.

In the weakly frustrated and unfrustrated regime $\lambda < \lambda_{\rm c}({\rm H},{\rm F}_{1/6})$, we have carried out the exact diagonalization for the system sizes up to $\Ns=18$ to obtain the phase diagrams of Figs. \ref{phase_stag} and  \ref{phase_stag_ferri}. The spontaneous magnetization obtained by numerical diagonalization is shown in Fig. \ref{fig:ms_stag}. 
We find no evidence supporting the presence of PF ground states in the thermodynamic limit within the numerical accuracy. 

\section{Ground-State Properties of the S1DC with Type B Distortion}

\subsection{Weak distortion regime}\label{subsec:weakb}
In the case of the type B distortion, the effective interaction between the spins of two cluster-$n$'s separated by a 
dimer 
 is ferromagnetic for small $\deltab$ 
as discussed in Ref. \citen{hts_distort}.
Therefore, we expect the ferrimagnetic ground state with spontaneous magnetization quantized as $m=1/(3(n+1))$ per site for small $\deltab$ in the range $\lambda_{\rm c}(n,n+1) < \lambda < \lambda_{\rm c}(n-1,n)$. Following Ref. \citen{hts_distort}, we call this phase a ferrimagnetic DC$n$ phase (FDC$n$ phase). 
In contrast, the ground state for $ \lambda_{\rm c}({\rm H},{\rm F}_{1/6}) <\lambda <\lambda_{\rm c}(3,{\rm H})$  remains in the Haldane phase, since a nonmagnetic gapped phase without STSB is generally robust against weak distortions. This phase is a symmetry protected topological phase with half-integer edge spins.
\subsection{Strong distortion regime $(\deltab \simeq 1)$}\label{subsec:strongb}

For $\deltab=1$ and $\lambda=0$, the whole system is decomposed into two parts. One is a single spin-1 chain of length $2N$ consisting of $\v{S}_l$ and $\v{\tau}^{(1)}_l$ with exchange constant $2J$ described by the Hamiltonian
\begin{align}
\H_0&=\sum_{i=1}^{2N}2J\v{\sigma}_l\v{\sigma}_{l+1}\label{eq:main_b}
\end{align}
where $\v{\sigma}_{2l}=\v{\tau}^{(1)}_l$ and $\v{\sigma}_{2l+1}=\v{S}_l$. The ground state of the Hamiltonian (\ref{eq:main_b})  is the nonmagnetic Haldane state $\ket{\rm H}$ with energy $E_{\rm H}$. 
The remaining part is $N$ isolated spins $\v{\tau}^{(2)}_l$.

For small $1-\deltab$ and $\lambda$, the spins $\v{\tau}^{(2)}_l$ interact with each other mediated by the fluctuation in the chain (\ref{eq:main_b}). Since the correlation within the Haldane chain (\ref{eq:main_b}) is short ranged, we consider only the nearest-neighbour effective coupling $\Jeffb$ between $\v{\tau}^{(2)}_l$ and $\v{\tau}^{(2)}_{l+1}$. Then, 
$\Jeffb$ can be estimated by the second order perturbation calculation as
\begin{align}
\Heffb&=\sum_{i=1}^N\Jeffb\v{\tau}^{(2)}_l\v{\tau}^{(2)}_{l+1}\\
\Jeffb&=2(F(0)+2F(2)+F(4))(1-\deltab)^2\nonumber\\
&+4(F(1)+F(3))(1-\deltab)\lambda\nonumber\\
&+2F(2)\lambda^2
\end{align}
where $F(l)$ is defined by  
\begin{align}
F(l)=-\sum_{\alpha}\frac{\bra{\rm H}\sigma_i^z\ket{\alpha}\bra{\alpha}\sigma_{i+l}^z\ket{\rm H}}{E_{\alpha}-E_{\rm H}}
\end{align} 
Here, $\ket{\alpha}$ and $E_{\alpha}$ are the eigenstate and eigenenergy of the $\alpha$th excited state of $\H_0$. We estimated the values of $F(l)$ ($l=1,...,4$) for the finite length spin-1 chain (\ref{eq:main_b}) with $2N=8, 10$ and 12. After the extrapolation to $N\rightarrow\infty$, we find that $\Jeffb$ is positive for
\begin{align}
0.594 \gtrsim \frac{1-\deltab}{\lambda} \gtrsim 0.416 \label{eq:nmag_cond_b}
\end{align}
and negative otherwise. Hence, the ground state of the chain consisting of spins $\v{\tau}_l^{(2)}$ is a nonmagnetic Haldane state in the region (\ref{eq:nmag_cond_b}) and a ferrimagnetic state with total magnetization $M=N$ otherwise. As a whole diamond chain, the former corresponds to the double Haldane phase\cite{hiki1,hiki2,kol1,kol2,kol3} and the latter to the ferromagnetic phase with $m=1/3$. The double Haldane phase consists of two coupled chains with Haldane ground state. In contrast to the previously studied examples of double Haldane phases that consist of two Haldane chains with equal length, the length of  one Haldane chain $H_0$ is two times larger than the other one $\Heffb$ in the present case.  Nevertheless, this ground state is topologically trivial, since the the edge spins of two Haldane chains can locally cancel out. The phase boundary (\ref{eq:nmag_cond_b}) is consistent with the numerical phase diagram presented in the next section around $(\lambda,\deltab)=(0,1)$. 
\subsection{Numerical phase diagram}
\begin{figure} 
%\centerline{\includegraphics[width=6.5cm]{ferri_uni_n18.eps}}
\centerline{\includegraphics[width=6.5cm]{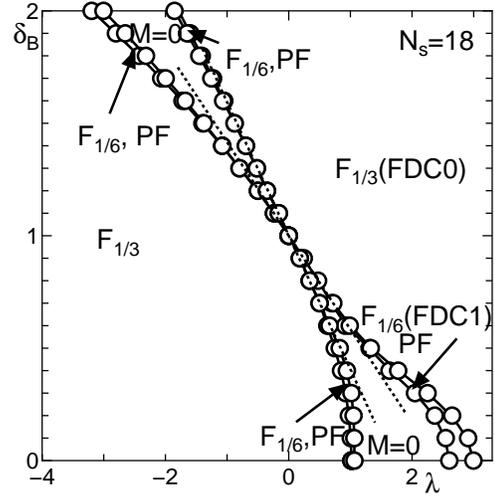}}
\caption{Phase diagram of the S1DC with type B distortion with $\Ns=18$. 
 The dotted lines are the boundaries determined by Eq. (\ref{eq:nmag_cond_b})}
\label{phase_ferri}
\end{figure}

For finite $\deltab$, we determined the ground-state phase diagram  
by the numerical diagonalization for the system size $\Ns=18$, as shown in Fig.~\ref{phase_ferri}. 
Among system sizes tractable by  numerical diagonalization, only the size $\Ns=18$ is compatible with all the ground-state structures with $n=0, 1$, and 2. 
As expected, the FDC$n$ quantized ferrimagnetic phases with $m=1/(3(n+1))$ are found for these values of $n$.

%=====================================
%=====================================
%=====================================
\begin{figure} 
%\centerline{\includegraphics[width=7cm]{mag040sel.eps}}
\centerline{\includegraphics[width=7cm]{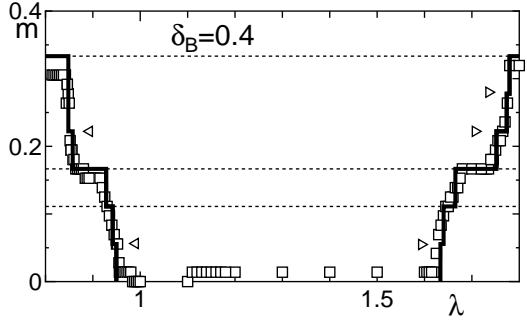}}
\caption{Spontaneous magnetization for $\deltab=0.4$. The exact diagonalization results for $\Ns=18$ with periodic boundary condition are shown by thick solid lines and DMRG results for $\Ns=72$ with open boundary condition are shown by the open squares. The dotted lines indicate the values of the spontaneous magnetization $m=1/3$, 1/6 and 1/9 
in the FDC$n$ phases. The left and right triangles indicate the positions of other steps for $\Ns = 18$ that suggest the possibility of PF phases.}
\label{mag_typeb}
\end{figure}
%=====================================

By inspecting numerical data for $\Ns=18$, 
we also find  narrow steps where the spontaneous magnetization does not satisfy $m=1/(3(n+1))$ for any integer $n$ between quantized ferrimagnetic phases as shown in Fig. \ref{mag_typeb} by thick solid lines for $\deltab=0.4$. The positions of these steps are indicated by the left and right triangles. These steps suggest the presence of PF phases. The ferrimagnetic phase of this kind has been found in various frustrated one-dimensional quantum spin systems.\cite{furuya,sekiguchi}
The DMRG calculation for larger $\Ns$ also supports the presence of PF phases in addition to the quantized ferrimagnetic phase with $m=1/1/3$ and $1/6$ as shown in Fig. \ref{mag_typeb} by open squares for $\deltab=0.4$.
 The origin of these PF phases can be understood by the same argument as that for the mixed diamond chain with $(S,\tau)=(1,1/2)$.\cite{hts_distort} In contrast to the case of type A distortion, PF phases are also present for $\lambda < \lambda_{\rm c}({\rm H}, {\rm F}_{1/6})$.  The physical mechanism to stabilize the latter PF phases remains unresolved.

\begin{figure} 
\centerline
%{\includegraphics[height=1.5cm]{lattice_s1dia_c_uni_bd_mono.eps}}
{\includegraphics[height=1.5cm]{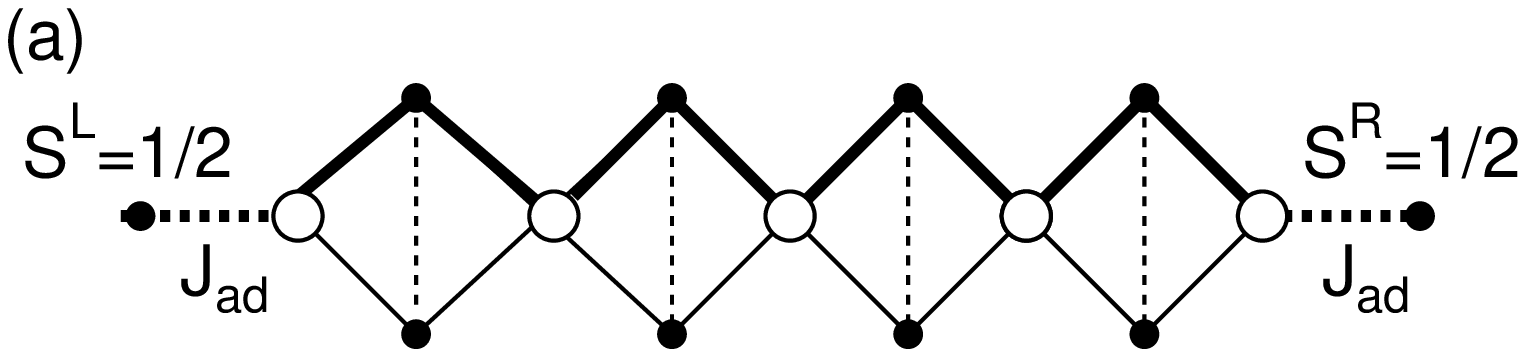}}
\centerline
%{\includegraphics[height=1.5cm]{lattice_s1dia_c_uni_bd_mono.eps}}
{\includegraphics[height=1.5cm]{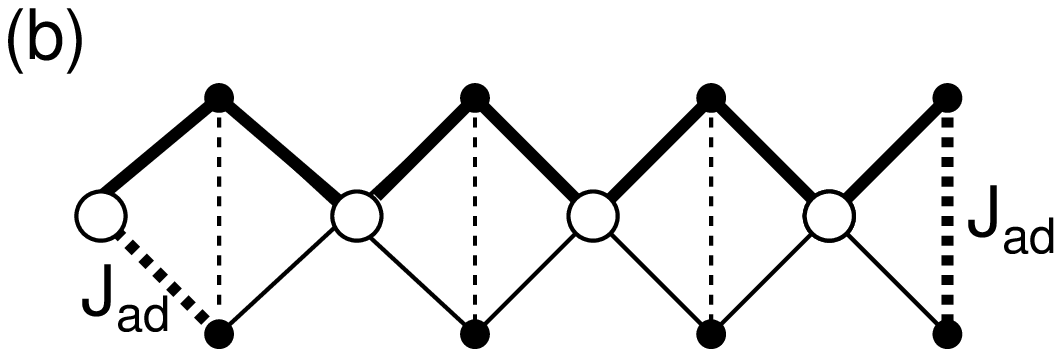}}
\caption{
Structures (a) H and (b) DH used for the calculation of the energy gaps $\Delta_{\rm H}$ and  $\Delta_{\rm DH}$ in the nonmagnetic phase, respectively.} 
\label{fig:modified_lattice}
\end{figure}

As discussed in the subsections \ref{subsec:weakb} and \ref{subsec:strongb}, a Haldane phase, which is a symmetry protected topological state, and the trivial double Haldane phase are present in the nonmagnetic phase. 
 To distinguish these two phases, we carry out the finite size DMRG calculation and estimate the energy gap with structures H and DH depicted in the Figs. \ref{fig:modified_lattice}(a) and (b), respectively. The number of states $\chi$ kept in each subsystem ranged from 480 to 840. For the structure H, additional spins $\v{S}^{L}$ and $\v{S}^{R}$ with magnitude 1/2 are added to both ends with exchange coupling  $\Jad (\v{S}^{L}\v{S}_{1}+\v{S}^{R}\v{S}_{N})$ to compensate the edge spins with magnitude 1/2 at both ends of the open Haldane chain. Then, the energy gap $\Delta_{\rm H}$ in the Haldane phase should be finite in the thermodynamic limit. On the contrary, in the structure DH, the interaction $(1-\deltab)\v{S}_{1}\v{\tau}_1^{(2)}+\lambda\v{\tau}_N^{(1)}\v{\tau}_N^{(2)}$  is replaced by $\Jad(\v{S}_{1}\v{\tau}_1^{(2)}+\v{\tau}_N^{(1)}\v{\tau}_N^{(2)})$ so that the edge spins of two Haldane chains are coupled antiferromagnetically to form a nonmagnetic singlet pair on each end.  Then, the energy gap  $\Delta_{\rm DH}$ in the double Haldane phase should be finite.
%=====================================

\begin{figure}

\centering

\includegraphics[width=7cm]{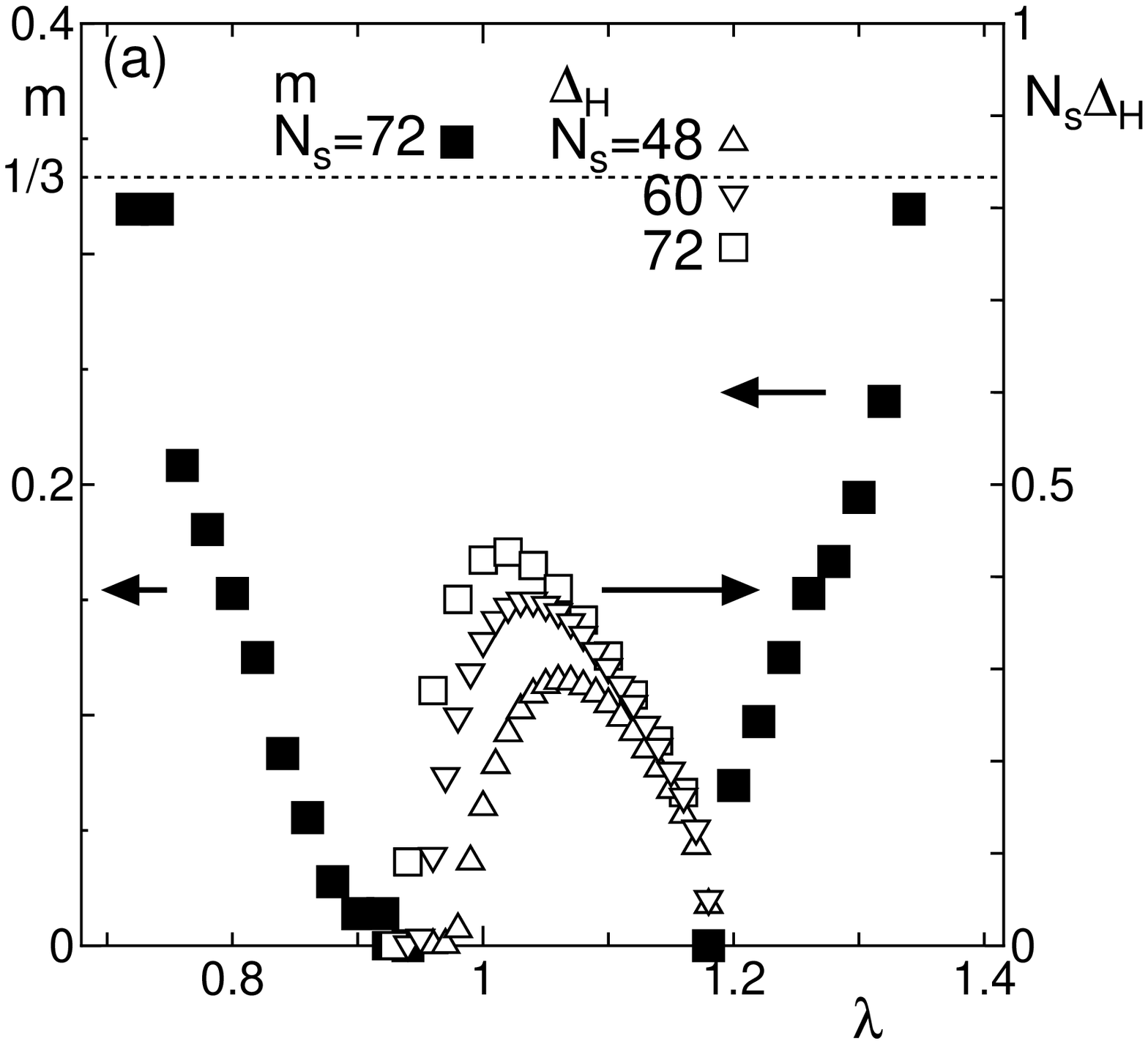}

\includegraphics[width=7cm]{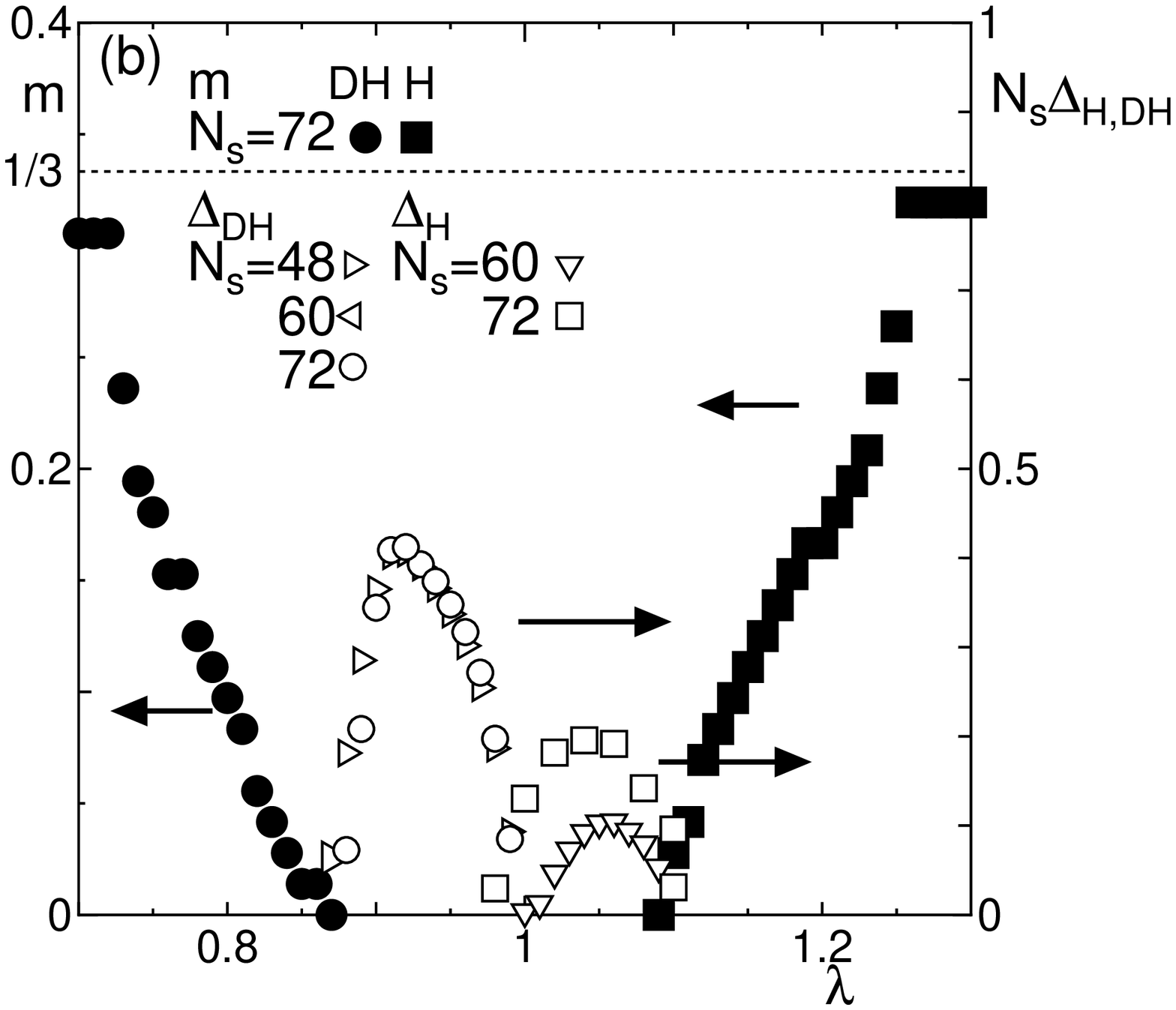}

\includegraphics[width=7cm]{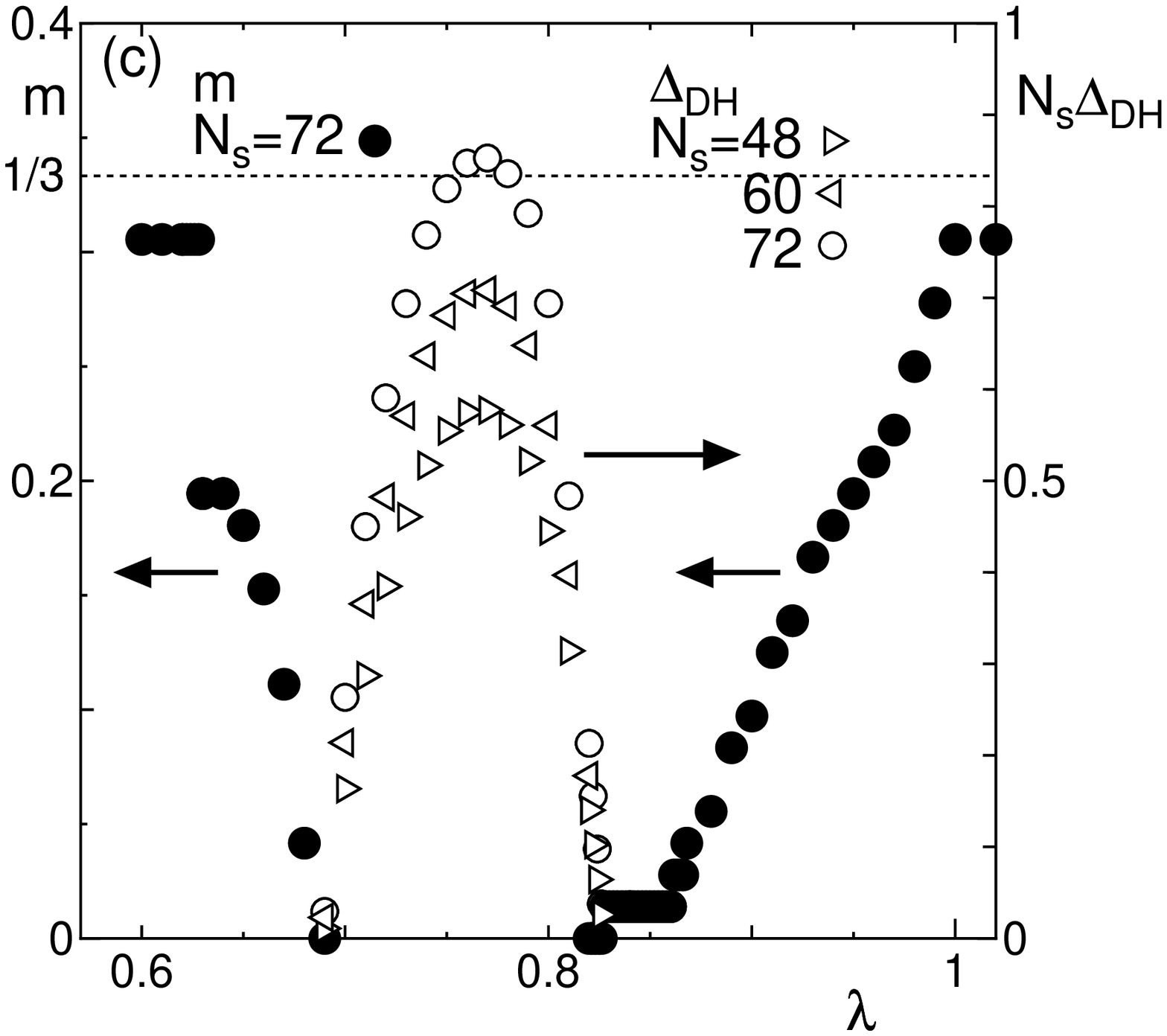}
\caption{Spontaneous magnetization $m$, scaled energy gaps $\Ns\Delta_{\rm H}$ and $\Ns\Delta_{\rm DH}$ for (a) $\deltab=0.5$,  (b) $\deltab=0.52$ and  (c) $\deltab=0.6$. The spontaneous magnetization for structure H (DH) is shown by filled squares (circles). }
\label{fig:gap}
\end{figure}

The numerical results for $\deltab=0.5$, 0.52 and 0.6 are shown in Fig. \ref{fig:gap}(a), (b) and (c), respectively. We set the coupling $\Jad=1$. In the nonmagnetic phase, the scaled gaps  $\Ns\Delta_{\rm H}$ and $\Ns\Delta_{\rm DH}$ are shown for several values of $\Ns$ by open symbols. Here, $N_{\rm s}$ is the number of spins including the additional spins. In the ferrimagnetic phase, the spontaneous magnetization per spin $m$ is plotted for $\Ns=72$. The quantized ferrimagnetic phase is so narrow  that it is numerically undetectable in this regime. 

For $\deltab=0.5$, the whole nonmagnetic phase belongs to the Haldane phase since the scaled gap $\Ns \Delta_{\rm H}$ increases with the system size as shown in Fig. \ref{fig:gap}(a).  For $\deltab=0.52$, the transition between the Haldane and double Haldane phase takes place as shown in Fig. \ref{fig:gap}(b). Unfortunately, even in the region where the finite size energy gap $\Delta_{\rm DH}$ is finite, the scaled gap $\Ns\Delta_{\rm DH}$ is almost independent of the system size. This means that the correlation length is larger than the length of the system employed in the numerical analysis. Considering the continuity to larger values of $\deltab$, we expect this region is in the double Haldane phase with large correlation length. Actually, for $\deltab=0.53$, we clearly found that $\Ns\Delta_{\rm DH}$ increases with the system size in the corresponding region. Nevertheless, we have chosen to present the data for $\deltab=0.52$, since $\Delta_{\rm H}$ in the Haldane phase is numerically undetectably small for  $\deltab=0.53$.

 For $\deltab=0.6$,  the scaled gap $\Ns \Delta_{\rm DH}$ increases with the system size as shown in Fig. \ref{fig:gap}(c) for $0.69 \lesssim \lambda \lesssim 0.826$. Hence, this region clearly belongs to the double Haldane phase. The spontaneous magnetization $m$ is clearly finite for $\lambda \gtrsim  0.86$ and $\lambda \lesssim  0.69$. For $0.826 \lesssim \lambda \lesssim  0.86$, however, the ground state has magnetization $M=1$ for $\Ns=72$. It is not clear whether this spontaneous magnetization remains finite in the thermodynamic limit. Hence, we cannot rule out the possibility of a new nonmagnetic phase in this region.

\section{Summary and Discussion}

We investigated the ground-state phases of S1DC with two types of distortion, type A and type B.
In the region where the ground states of the undistorted S1DC are DC$n$ states with finite $n$,  the effective interaction between the cluster spins is antiferromagnetic for the type A distortion and ferromagnetic for the type B distortion. 
Hence, for the type A distortion, the DC$n$ ground states are transformed into the HDC$n$ ground states. 
The nature of the HDC$n$ phase is essentially the same as that of the mixed diamond chain with $(S,\tau)=(1,1/2)$. Hence, we have determined the phase diagram  in the same way as in Ref. \citen{hts_distort}. For the type B distortion, the DC$n$ ground states are transformed into the ferrimagnetic FDC$n$ ground states. In addition to the FDC$n$ phases with quantized spontaneous magnetization $m=1/(3(n+1))$, 
the PF phases are also found numerically between the FDC$n$ and FDC$(n+1)$ phases.

If the ground state of the undistorted S1DC is the uniform Haldane or ferrimagnetic F$_{1/3}$ or F$_{1/6}$ phases, they are robust against a weak distortion. The quantized ferrimagnetic phases, however, shift to the small $\lambda$ regime with the increase of distortion. For the type B distortion, the PF phases emerge between the quantized ferrimagnetic phases, while we found no such evidence for the type A distortion within our numerical calculation. The physical origin of this difference is left for future studies.
 
For the type B distortion, a nonmagnetic region is present in the intermediate frustration regime between two types of ferrimagnetic phases. In this region, two topologically distinct phases are identified, namely the Haldane phase, which is a symmetry protected topological phase, and the double Haldane phase, which is a trivial phase. The latter consists of two coupled Haldane chains. In contrast to the previously known examples of double Haldane phases for frustrated spin-1 Heisenberg chains that consists of two Haldane chains with equal length,\cite{hiki1,hiki2,kol1,kol2,kol3} the present double Haldane phase consists of Haldane chains with lengths $N$ and $2N$. Further investigation is required to elucidate the nature of this new type of double Haldane phase. The possibility of a nonmagnetic phase different from  these two phases is also suggested. Within our numerical data, however, it is not possible to conclude whether this is an artifact of the finite size calculation or remains in the thermodynamic limit. The investigation of these states is left for future studies.

In contrast to the distorted mixed diamond chain with $(S,\tau)=(1,1/2)$, which has no experimental counterpart so far, the S1DC with type A distortion is already realized as experimental material.\cite{kuni,kiku1} Although the ground state of this material is ferrimagnetic, the realization of the materials with other exotic ground states such as the Haldane phases with STSB or the double Haldane phase would be hopefully within the scope of experimental studies in the near future.

The author thanks K. Takano, K. Okunishi, and T. Hikihara for valuable discussion and comments. The numerical diagonalization program is based on the package TITPACK ver.2 coded by H. Nishimori.  Part of the numerical computation in this work has been carried out using the facilities of the Supercomputer Center, Institute for Solid State Physics, University of Tokyo,  and the  Yukawa Institute Computer Facility, Kyoto University. 
This works is  supported by  JSPS KAKENHI Grant Number JP25400389.

\end{document}